\documentclass[runningheads]{llncs}
\usepackage{graphicx}
\usepackage{subfig}
\usepackage[ruled,vlined]{algorithm2e}
\newcommand{\Sref}[1]{\S\ref{#1}}
\newcommand{\fref}[1]{figure~\ref{#1}}

\newcommand{\tref}[1]{table~\ref{#1}}

\begin{document}
\title{Polarizing Tweets on Climate Change\thanks{The authors would like to acknowledge the support of Department of Engineering and Public Policy, Carnegie Mellon University and Center for Computational Analysis of Social and Organizational Systems (CASOS), Carnegie Mellon University.}}
\titlerunning{Polarizing Tweets on Climate Change}
%
\author{Aman Tyagi\inst{1}\orcidID{0000-0002-6654-0670} \and
Matthew Babcock\inst{2}\orcidID{0000-0002-7228-0262} \and
Kathleen M. Carley\inst{1,2}\orcidID{0000-0002-6356-0238} \and
Douglas C. Sicker\inst{1,2}\orcidID{0000-0003-0296-5212}}
\authorrunning{Tyagi et al.}
%
\institute{Engineering and Public Policy, Carnegie Mellon University, PA 15213, USA \and
Institute for Software Research, Carnegie Mellon University, PA 15213, USA
\email{amantyagi@cmu.edu,mbabcock@andrew.cmu.edu, kathleen.carley@cs.cmu.edu,sicker@cmu.edu}}

\maketitle              

\begin{abstract}
We introduce a framework to analyzes the conversation between two competing groups of Twitter users, one who believe in the anthropogenic causes of climate change (Believers) and a second who are skeptical (Disbelievers). As a case study, we use Climate Change related tweets during the United Nation’s (UN) Climate Change Conference – COP24 (2018), Katowice, Poland. We find that both Disbelievers and Believers talk within their group more than with the other group; this is more so the case for Disbelievers than for Believers. The Disbeliever messages focused more on attacking those personalities that believe in the anthropogenic causes of climate change. On the other hand, Believer messages focused on calls to combat climate change. We find that in both Disbelievers and Believers bot-like accounts were equally active and that unlike Believers, Disbelievers get their news from a concentrated number of news sources.

\end{abstract}
\section{Introduction}
Social media platforms such as Twitter have become an important medium for debating and organizing around complex social issues \cite{climateSandB}. One such complex issue with significant socio-economic and political implications is climate change. Debates over climate change involve different groups with different inherent motivations and beliefs. For instance, among the people who are skeptical of climate science findings are people who outright reject the data that climate change is occurring, and others who argue that climate change is occurring due to non-anthropogenic causes. Similarly, there is significant difference in beliefs among people who believe in anthropogenic causes of climate change. Work by \cite{mathewspaul} argue that there are groups who believe that impact of climate change is exaggerated (so-called “luke-warmers”), others who argue that we need an across-the-board technological change in energy production\cite{shellenberger_nordhaus_2004} or even end of capitalism\cite{klein2015changes,angus2016facing}. Furthermore, some groups argue that it is already too late to avoid climate catastrophe\cite{scranton2015learning}. In this paper, we analyze conversations between two broad competing groups of Twitter users, one who believes in anthropogenic causes of climate change (Believers) and a second who are skeptical or outright deny climate change is occurring (Disbelievers). To this end, we classify users into Disbelievers and Believers in Twitter conversations.

In this paper, we present a case study by analyzing conversations- Believers and Disbelievers, during United Nation’s (UN) Climate Change Conference – COP24 (2018), Katowice, Poland. Previous studies about climate change discussions on social media, such as \cite{WILLIAMS2015126} and \cite{cody2015climate}, lacked the context of a significant event. They also didn’t take into account the behavior of bots during such an event. We examine what role, if any, that bots play within Disbeliever and Believer competing groups. By restricting the study to a particular event, we were able to manually inspect large fractions of stories in the competing groups. This case study should be helpful to inform future studies regarding climate change conversations on social media that cover longer time span. Our research questions are as follows:
\begin{enumerate}
    \item What are the conversational subtopics within the Believer and Disbeliever groups, and what does common word use by these competing groups highlight? Do individuals of one group interact with individuals of the other group? What are the popular sources of information within these groups?
    \item Are bots more active in one particular group over another?
\end{enumerate}
We analyze these research questions using Twitter conversations on climate change during COP24. We describe our data collection method in \Sref{sec:data-collect}. We use hashtag based method to classify users into Disbelievers and Believers described in \Sref{sec:method}. In \Sref{sec:results} and \Sref{sec:discussion} we present our results and their implications. Through this research study we provide a framework to analyze polarizing networks and the implications for climate change discussion.

\section{Data Collection and Method}
\label{sec:data-method}
Twitter has been an important social media platform to study conversations about natural disasters, medical decisions, race relations and numerous other important issues \cite{dredze2017vaccine}. We look at four main types of communication on Twitter in this paper: 1) Tweeting, 2) Retweeting, 3) Replying, and 4) Mentioning. We call the sum of the four types of communication as “all communication”. In this paper, we look at these communications as networks and find network measures to compare and contrast communication from and between Disbelievers and Believers. 

\subsection{Data Collection}
\label{sec:data-collect}
UN Framework Convention on Climate Change’s (UNFCCC) Conference of Par-ties (COP) is an annual meeting of different states represented at the UN and acts as a venue to discuss the progress and establish obligations with regards to responding to climate change \cite{unitednationscop}. This event provided an opportunity to look at the Disbeliever and Believer climate change messaging on Twitter in context of a significant event. 

We collected tweets with hashtags and certain keywords from November 27th to December 20th, 2018 using Twitter’s API. We decided on collection hashtags based on hashtags related to \#climatechange found on best-hashtags.com. We added more keywords based on these hashtags and news articles found after searching for keyword “COP24” on Google \footnote{Hashtags and keywords used for collection:\#COP24, \#ClimateChange, \#ClimateHoax, \#ParisAgreement,\#IPCC, \#InsideCOP24,\#Climate, \#ClimateChangeisReal, \#ClimateAction, \#GlobalWarming, COP24, Climate Change, Paris Agreement, Climate Hoax, IPCC, Climate, Global Warming}. The combined data set contains a total of 1,379,584 distinct tweets (including retweets).

\subsection{Method}
\label{sec:method}
We identified competing groups of Believers and Disbelievers by hashtags used by these groups. Hashtags have been shown to be a realistic substitute to identify stances among different groups on social media \cite{evans2016stance}. For example, previous studies suggest that climate Disbelievers use terms such as hoax and scam \cite{runciman_2015}. We analyzed common hashtags used in our dataset and found that “ClimateHoax” and “ClimateChangeIsReal” hashtags are used mostly by Disbelievers and Believers respectively. There are 528 distinct tweets with keyword “\#ClimateHoax” and 9,008 tweets with keyword “\#ClimateChangeIsReal” in our data set. We manually checked all tweets with hashtag “ClimateHoax” and randomly sampled 1,000 tweets from data subset with hashtag “ClimateChangeIsReal”. We identified ~96\% of tweets with “\#ClimateHoax” as climate change Disbeliever tweets. For “\#ClimateChangeIsReal” out of the 1,000 randomly selected tweets, we identified about 99\% as climate Believer tweets. We therefore conclude that hashtag “ClimateHoax” and hashtag “ClimateChangeIsReal” can be used as proxies for tweets broadcasted by Disbelievers and Believers respectively in our data set. 

To identify more hashtags used by Believers and Disbelievers, we use the method described in \cite{tyagi2020computational}. We choose hashtags which are most used with hashtag “ClimateHoax” and hashtag “ClimateChangeIsReal” and are associated with conspiracy in case of Disbelievers or have similar meaning to “ClimateChangeIsReal” in case of Believers \footnote{Disbeliever hashtags: ClimateHoax, YellowVests and Qanon. Believer hashtags: ClimateChangeIsReal,ClimateActionNow, FactsMatter, ScienceMatters, ScienceIsReal}. We give an initial weight of -1 to Disbeliever hashtags and +1 to Believer hashtags. We use these labels in a weighted hashtag x hashtag co-occurance network, to find an average label from -1 to 1 for other hashtags. The method used for propagating labels to other hashtags is reported in Algorithm 1. We aggregate hashtags used by each user and found a weighted average of all hashtags used by a particular user. We label a user as Disbeliever, Believer or unclassified if the weighted average was negative, positive or zero respectively. We assume that within our collection period Disbelivers or Believers do not change their stance and hence unlike in \cite{tyagi2020computational} we only look at aggregate polarized hashtags over entire dataset. Overall, we found a set of 8,413 tweets from 2,170 Disbelievers and 120,497 tweets from 15,640 Believers. We randomly sampled 100 users from both groups of users and manually checked their timeline to find approximately 91 percent of Disbelievers as showing activity akin to a Disbeliever and about 96 percent of Believers showing activity akin to a Believer.

\begin{algorithm}[ht]
\scriptsize
\SetAlgoLined
\KwIn{Graph \emph{G}; Nodes = \emph{n}; Edges = \emph{e}; Edge Weight =  $\emph{$e_{ij}$}$, $i \in n$ and $j \in n$}
 \textbf{initialize} $\gamma=100$ and \emph{i}=0\;
 \For{each n}{
  \textbf{define} $l$ = integer(i/$\gamma$); $i$+=1\;
  \For{each n}{
      \If{n not labeled}{
       \textbf{compute} $t$ = neighbors of $n$\;
       \textbf{compute} $t_l$ = labeled neighbors of $n$\;
       \If{$|t_l| + l \geq t$}{
        \textbf{initialize} \textit{score}, $c$\\
        \For{each $t_i \in t$}{
            score += label $t_i * e_{n{t_i}}$\\
            c += $e_{n{t_i}}$\\
            }
        \textbf{update} label $n = score/c$
       }
        }
    }
 }
 \caption{Label Propagation Algorithm}
\end{algorithm}

\paragraph{BOT Detection:}To find bots accounts in our data set, we used CMU’s Bot-Hunter \cite{beskow2018bot1,beskow2018bot2}. The output of Bot-Hunter is a probability measure of bot-like behavior assigned to each account. Unless otherwise stated, we report our analysis for a probability threshold of 0.5,as done in various machine learning classification methods \cite{pedregosa2011scikit}. In other words,we classified an account as bot-like if output probability from Bot-Hunter was greater than 0.5. At 0.5 threshold level we found 596,282 bot-like accounts out of total 1,035,416 users in our data set.

\paragraph{Account Type:} We used a classification model trained on the users' tweets and personal descriptions to find news agency accounts associated with our list of user accounts. The model is similar to the state-of-the-art model used in \cite{huangdetect2020}. The paper describes the model as a long-short term memory neural network \cite{hochreiter1997long} with an attention mechanism \cite{bahdanau2014neural}. In total, we find 2.2\% of Believer tweets as classified to be from news agencies and 6.2\% of Disbeliever tweets as classified to be from news agencies.

\section{Results}
\label{sec:results}
We begin by discussing topics of discussion within Believers and Disbelievers. Then we look at inter-group and intra-group interaction. Lastly, we look at the popular news agencies and contrast bot-like account’s behavior in these two groups. 

\subsection{Topics of Discussion}
To understand the conversations of both the gropus we found the most frequent words used by these competing groups. The results are presented below: \footnote{Note that in the construction of unigrams, we exclude common stop words.}\\ 
\textit{Believers:} climate, change, world, us, need, action, un, global, leaders, future\\
\textit{Disbelievers:} climate,change,global,private, us, un, sanders,world, end,warming

We find that Believers use words such as “need”, “action”, “leaders” and “future” more often, potentially indicating tweets calling for action to combat climate change. On the other hand, Disbelievers use words such as “private” (referencing "private jet"), “sanders”, “end” and “warming”, potentially indicating attacks on pro-climate change personalities and their messaging.

To further our understanding of the conversations and to find topics of opinions within these group, we performed topic modelling of tweets by Believers and Disbelievers using Latent Dirichlet Allocation (LDA) \cite{blei2003latent}. We ran our model to find top ten topics on the unigrams of tweets generated after removing common stop words. Among the top ten topics we report the top 3 list of words we were able to infer topics about in \tref{tab:table4}. In the first topic Disbelievers use words such as “scam” and “fakenews” with words associated with “climate change”, potentially calling out climate change as scam or fake. In the second topic Disbelievers are calling out personalities believing in human caused climate change.  In the third topic Disbelievers are talking about yellowvests movement which relates to the French movement against raising fuel taxes based on climate policy \cite{cigainero_2018}. On the other hand, for the first topic Believers use words related to using renewables and giving up fossil fuel. This can be inferred from the use of word “keepitintheground”, as the word is used on social media to ban any new use of fossil fuel \cite{chambercommerce_2017}. The second topic for Believers is about the climate change politics in Australia with words such as “auspol” (short for Australian politics) and “stopadani”. Specifically, “stopadani” is used in social media to protest against Adani group of companies digging Carmichael coal mine in Queensland, Australia \cite{zajac}. Lastly, the third topic for Believers relates to COP24 with word “takeyourseat” used in COP24 to signify the People’s seat initiative launched by the UN \cite{unitednationscop}.

\begin{table}
\footnotesize
\centering
\caption{Table of top 10 words (excluding hashtags) used by Disbelievers and Believers.}
\begin{tabular}{|c|c|c|c|c|c|}
\hline
 \multicolumn{3}{|c|}{\textbf{Disbelievers}}  & \multicolumn{3}{|c|}{\textbf{Believers}}  \\
\hline
climate	& climate	& yellowvests	& science & climatechangeisreal & cop \\
scam & bernie & maga & climate	& climatechange & climateaction \\
change & sanders & trump & year & auspol & climate \\
nuclear & travel & carbontax & keepitintheground & climatestrike & katowice \\
industry & potus & france & climateemergency & climateactionnow & world \\
fakenews & month & macron & record & greennewdeal & takeyourseat \\
crisis & change & policy & renewableenergy & climate & leader \\
record & planet & french & bcpoli & cdnpoli & solar \\ 
global & great & people & end & stopadani & change \\
agw & face & hoax & fact & globalwarming & poland \\
\hline
\end{tabular}
\label{tab:table4}
\end{table}

\subsection{User Accounts and Conversations}

\begin{figure}[htbp]
\footnotesize
\centering
\subfloat[Retweet Network]{\label{fig:a}\includegraphics[width=0.3\linewidth]{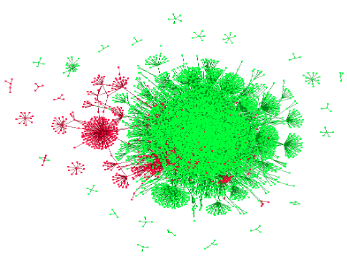}}\qquad
\subfloat[Mention Network]{\label{fig:b}\includegraphics[width=0.3\linewidth]{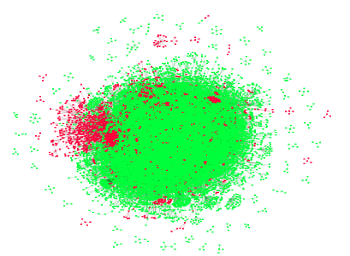}}\\
\subfloat[Reciprocal Network]{\label{fig:c}\includegraphics[width=0.3\textwidth]{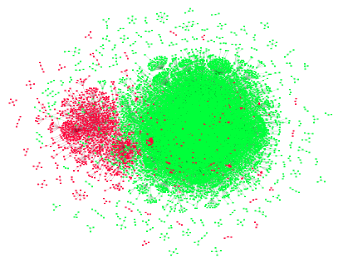}}\qquad%
\subfloat[Reply Network]{\label{fig:d}\includegraphics[width=0.3\textwidth]{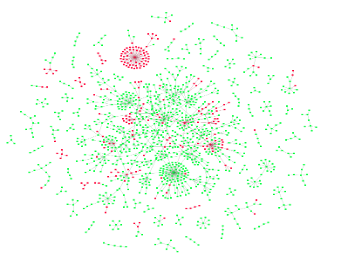}}%
\caption{Communication networks between Twitter accounts classified as Disbelievers (red) and Believers (green). The graphs were made using ORA-PRO \cite{carley2014ora,altman2017ora}}.
\label{fig:network}
\end{figure}

We first look at different Twitter networks formed from various communications to contrast Believers and Disbeliever. In \fref{fig:network}, we report figures for all four networks. In the retweet network, we can see a clear distinction between Believers and Disbelievers; Disbelievers retweet other Disbelievers more than they retweet Believers, and vice versa for Believers. The mentions network of Disbelievers and Believers shows more links between these groups meaning that Believers and Disbelievers do mention users from other groups on tweets. The reciprocal network has less activity between the groups than the mentions network, suggesting that although users from one group mention people from other group, they tend to have reciprocal relationships with their own group. The reply network has a much lower number of nodes compared to other networks which suggests that users in both groups prefer mentioning or retweeting rather than replying to tweets. The stark contrast in mentions and reciprocal activity confirms that users from one group do not engage in conversations with users from another group. After establishing differences in different type of behavior on Twitter, next we look at the combined communication of these groups to check how much these groups talk within themselves, i.e. how much “echo-chambery” these groups are. 

To compare echo-chamber effects in these two groups, we combine all the above networks to make a network of all communications to find echo-chamberness($e$)\footnote{For a unimodal network $G$, the $e$ of $G$ is $(r*d)^{(1/3)}$,where $r$ is the reciprocity of graph G, that is the fraction of edges in the graph that are reciprocal (a symmetric graph has $r = 1$), and $d$ is the density of graph $G$.} for each group with and without unclassified accounts. We find that for Disbelievers $e = 0.007$ and for Disbelievers with unclassified accounts $e = 0.003$. On the other hand, for Believers $e = 0.006$ and for Believers with unclassified accounts $e = 0.003$. The values of $e$ is small compared to a denser symmetric graph because the communications network does not represent the actual follower’s network of the users. The $e$ of both groups decreases on adding unclassified accounts, which indicates that each group is talking more to themselves, this is marginally truer for Disbelievers compared to Believers. 

The $e$ metric results suggests that more communication is happening within these groups compared to outside these groups. Next, we look at the difference in fraction of most crucial and influential spreaders of information in both the networks. This helps us determine whether or not these groups are influenced by multiple influencers or via a central actor. ORA-PRO twitter report labels users as “super spreader” as the most influential users in spreading information and “super friends” as most crucial users in bi-directional communication on twitter \cite{carley2014ora,altman2017ora}. Super spreaders in ORA-PRO are defined as user accounts in sum of mentioned-by and retweeted-by network which are in top 3 of following measures: 1) Often mentioned/retweeted by others, 2) Iteratively mentioned/retweeted by others, and 3) Often mentioned/retweeted by groups of others. To compare the two groups, we look at the fraction of user accounts labelled as super spreaders and super friends. We find that Disbelievers (0.48\%) have fractionally higher percentage of super spreaders than Believers (0.37\%). Disbelievers also have higher fraction of users classified as super friends than compared to Believers (0.38\% vs 0.28\%). We conclude that Disbelievers have higher fraction of influential users in the network compared to Believers.

\begin{figure}[htbp]
\footnotesize
\centering
{\label{fig:a1}\includegraphics[width=0.4\linewidth]{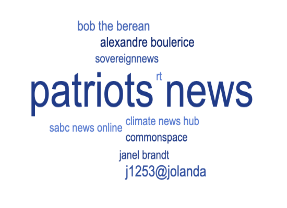}}\qquad
{\label{fig:b1}\includegraphics[width=0.4\linewidth]{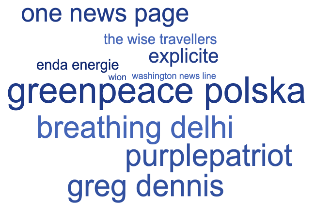}}\\
\caption{Word cloud of tweets by news agencies classified as Disbeliever (left) and Believer (right).}
\label{fig:news}
\vspace{-3mm}
\end{figure}

We look at the popular news sources within our different groups. In \fref{fig:news}, we present a word cloud of the names of accounts classified as news agency (\Sref{sec:method}) by the number of tweets in the competing groups. “Patriot News” dominates Disbelievers’ tweets (including retweets), but for Believers there is no one account which dominates.

\begin{figure}[htbp]
\footnotesize
\centering
\includegraphics[width=0.6\linewidth]{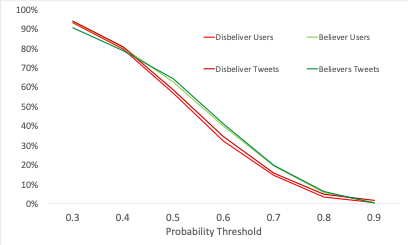}
\caption{Percentage of bots and tweets at different probability threshold for an account to be classified to be bot-like as predicted by Bot-Hunter \cite{beskow2018bot1} for climate Disbelievers and Believers group.}
\label{fig:bots}
\vspace{-3mm}
\end{figure}

Next, we compare bot-like activity in the two groups of Believers and Disbelievers. In \fref{fig:bots}, we report the bot-like account’s activity at different probability thresholds for an account to be classified to be bot-like for the Believers and Disbelievers. We find that the fraction of tweets and user accounts classified as bots are similar for both the groups at all threshold levels. This indicates that bots are similarly active in both the groups.

\section{Discussion}
\label{sec:discussion}
We classify users into two competing communities to compare and contrast the hashtags, bot percentage and messaging in the communities. We use climate change as case study to find groups with opposite views. An important finding of this paper is that different communities in climate change discussion primarily use different sets of hashtags. We find that Disbelievers words usage focus more on attacking personalities believing in anthropogenic origin of climate change and their messaging; on the other hand, Believers words usage focuses on callings to combat climate change. Our results indicate that unlike conversations on personalities in opposing groups, messages about social movements are dominated by discussion on movements aligned towards group’s beliefs rather than calling out movements driven by contrasting beliefs. 

We looked at the network structure of Believers and Disbelievers for these twitter interactions. We found greater homophily in retweet, reply and reciprocal networks compared to the mention network. This is consistent with the fact that typically users retweet, equivalent to resharing, if they endorse that message and are hence more likely to endorse a message from users aligned to their own perspective. On the other hand, mentioning activity could be a way to call out or malign members of other community. For all the types of networks, we found that both Disbelievers and Believers talk within their group more than with the other group; this is more so the case for Disbelievers than for Believers. Our results confirm findings from \cite{WILLIAMS2015126}, which concluded that there are segregated communities in climate change conversations on Twitter. Moreover, Disbelievers communication activity is influenced by higher fraction of users in their group compared to Believers. This coupled with the fact that Disbelievers are more “echo-chambery” suggests that higher fraction of conversations within Disbelievers happen with the influencers compared to Believers’ network. We conclude that Disbelievers are more organized around certain influencers in their network compared to Believers.

We found that unlike Believers, Disbelievers get their news from a concentrated number of news sources and hence may be more vulnerable to manipulation. We also found that in both Disbelievers and Believers bot-like accounts were equally active. This is in similar vein with previous findings that bot-like accounts tend to stir conversations in differently politically aligned groups rather than concentrating on conversations in one group \cite{bessi2016social}. We conclude that bot activity is further creating and nourishing the divide between Believers and Disbelievers. 

\bibliography{references.bib}
\bibliographystyle{splncs04}

\end{document}